\documentclass[aps,twoside,twocolumn]{revtex4}
\usepackage{graphicx}

\begin{document}

\title[]{Scaling Properties of Price Changes for Korean Stock Indices}

\author{Kyoung Eun \surname{Lee}}
%\email{jaewlee@inha.ac.kr}
%\affiliation{Department of Physics, Inha University Incheon 402-751 Korea}

\author{Jae Woo \surname{Lee}}
\email{jaewlee@inha.ac.kr}
\affiliation{Department of Physics, Inha University Incheon 402-751 Korea}

%\received{}
\date{\today}

\begin{abstract}
 We consider returns of two Korean stock market indices, KOSPI
and KOSDAQ index. Central
parts of the probability distribution function of returns
are well fitted by the Lorentzian distribution function. However,
tail parts of the probability distribution function follow a power law behavior well.
We found that the probability distribution function of returns for both
KOSPI and KOSDAQ, is outside the L\'{e}vy stable distribution.
\end{abstract}



\maketitle

\section{INTRODUCTION}

Complex behaviors of econophysics have greatly led to attentions
in the field of statistical physics. A lot of economic data have
been reanalyzed by  physicists recently\cite{MA99,MA97,MS95,BS94}.
Time series of stock market around the world have rich behaviors.
The time series deviate from the EMH(efficient market hypethesis).
Indices of stock market show scaling behaviors in the well
developed market. It is very difficult to understand the dynamics
of financial systems because there are many factors among
interacting agents.

\begin{figure}
\caption[0]{
The normalized return versus time for KOSPI
}
\end{figure}
\begin{figure}
\caption[0]{
The normalized return versus time  for KOSDAQ.
}
\end{figure}

\begin{figure}
\includegraphics[width=6cm,height=7cm,angle=270]{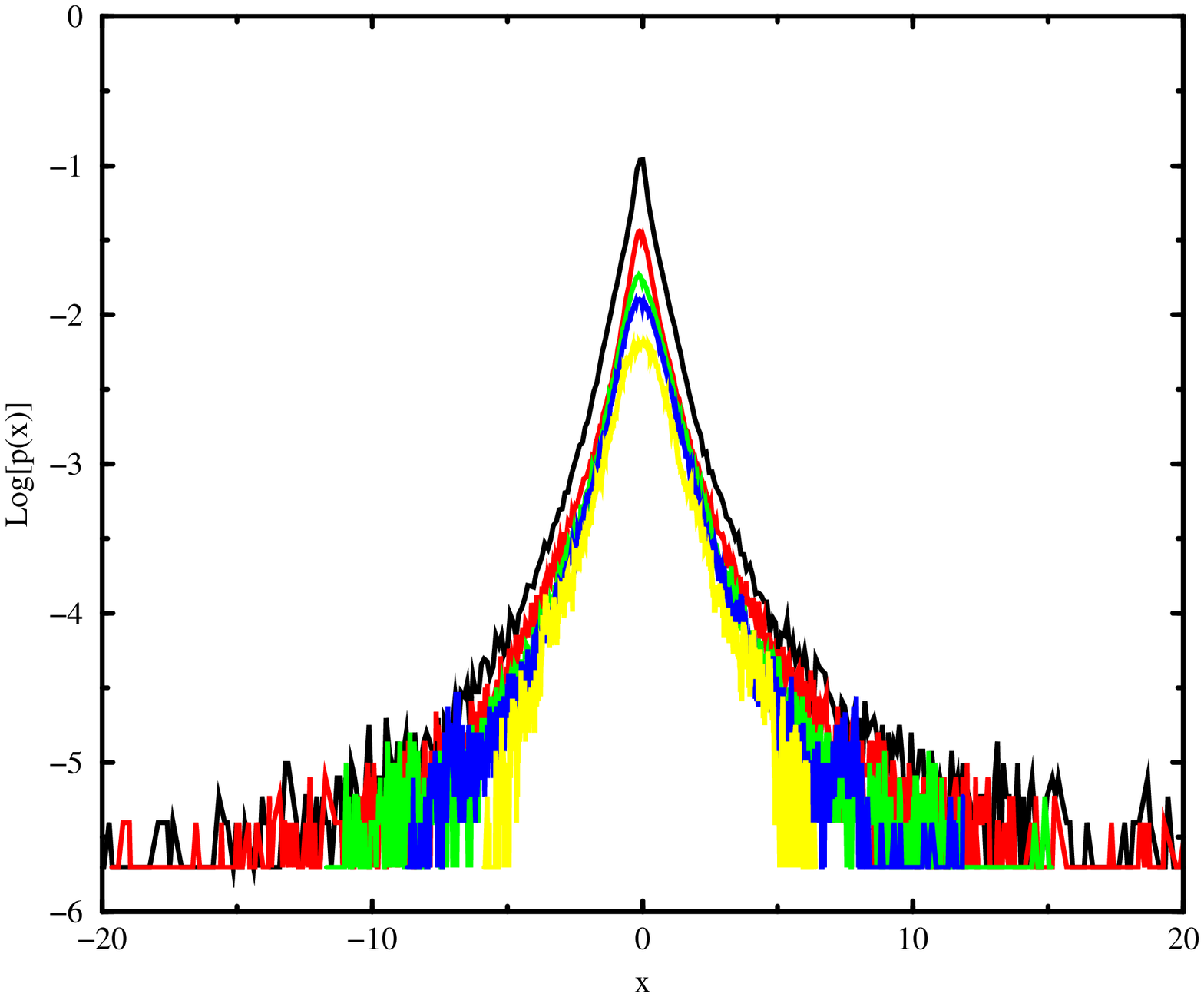}
\caption[0]{
The semilogarithmic plot of the probability distribution function
as a function of the return with
the return time T=1min(top), 10min, 30min, 60min and 600min(bottom) for KOSPI.
}
\end{figure}
\begin{figure}
\includegraphics[width=6cm,height=7cm,angle=270]{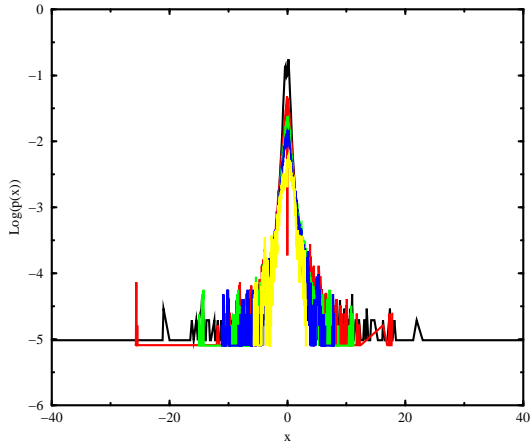}
\caption[0]{
The semilogarithmic plot of the probability distribution function
as a function of the return with
the return time T=1min(top), 10min, 30min, 60min and 600min(bottom) for KOSDAQ.
}
\end{figure}

Bachelier has proposed a financial model of stochastic process of
returns which consider the variation of share prices as an
independently, identically distributed (i.i.d) Gaussian random
variable\cite{BA1900}. However, the distribution of returns in
financial markets does not follow Gaussian distribution.
Mandelbrot analyzed a relatively short time series of cotton
prices and observed that returns have L\'{e}vy stable symmetric
distribution with Pareto fat tail\cite{MA63,FA63,LE,PA}.
 Boston group reported departures
from L\'{e}vy stable distribution of returns by analyzing high frequency data points
of the S \& P 500
index\cite{GP99,GM99,LK99,GP00,SA00,LH03,CL01,LL02,KC01,KP03,KK02,NA02}.
 They observed that large events was very frequent in the
data, a fact largely underestimated by a Gaussian process. They
also found a power-law behavior of the probability density
function (pdf) of returns with the fat tail exceeding the L\'{e}vy
stable distribution. The similar behavior reported for
distribution of returns of other indices, including
FOREX\cite{EV95}, DAX\cite{GD02}, and Hang-Seng indices\cite{WH01}

In this article we consider two Korean stock market indices, KOSPI
and KOSDAQ index. We consider a set of data recorded per one
second for KOSPI from March 30 1992 to November 30 1999 and for
KOSDAQ from  March 5 2001 to February 28 2003. KOSDAQ index is
recorded in 30 seconds interval. We count the time during trading
hours and remove closing hours, weekends and holidays from data
sets. For a time series $Z(t)$ of stock market index values, the
return $G_T (t)$ over a return time $T$ is defined as
\begin{equation}
G_{T} (t) = \ln{\frac{Z(t+T)}{Z(t)}}
\end{equation}
For small changes in $Z(t)$, the return is approximately
\begin{equation}
G_T (t) \simeq \frac{Z(t+T) -Z(t)}{Z(t)}
\end{equation}
The normalized return is defined as
\begin{equation}
g_T (t) = \frac{G_T (t) - < G_T (t) > }{\sigma(G_T (t))}
\end{equation}
where $\sigma(G)$ is the standard deviation and $< \cdots > $ denotes averaging
over time variable.

\begin{figure}
\includegraphics[width=6cm,height=7cm,angle=270]{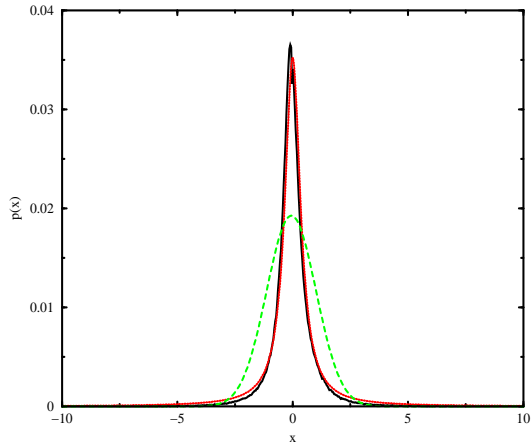}
\caption[0]{
Fitting of the probability distribution function
with the return time T=10min
by the Gaussian function (dashed line) and
by the Lorentzian (dotted line) function for  KOSPI.
}
\end{figure}
\begin{figure}
\includegraphics[width=6cm,height=7cm,angle=270]{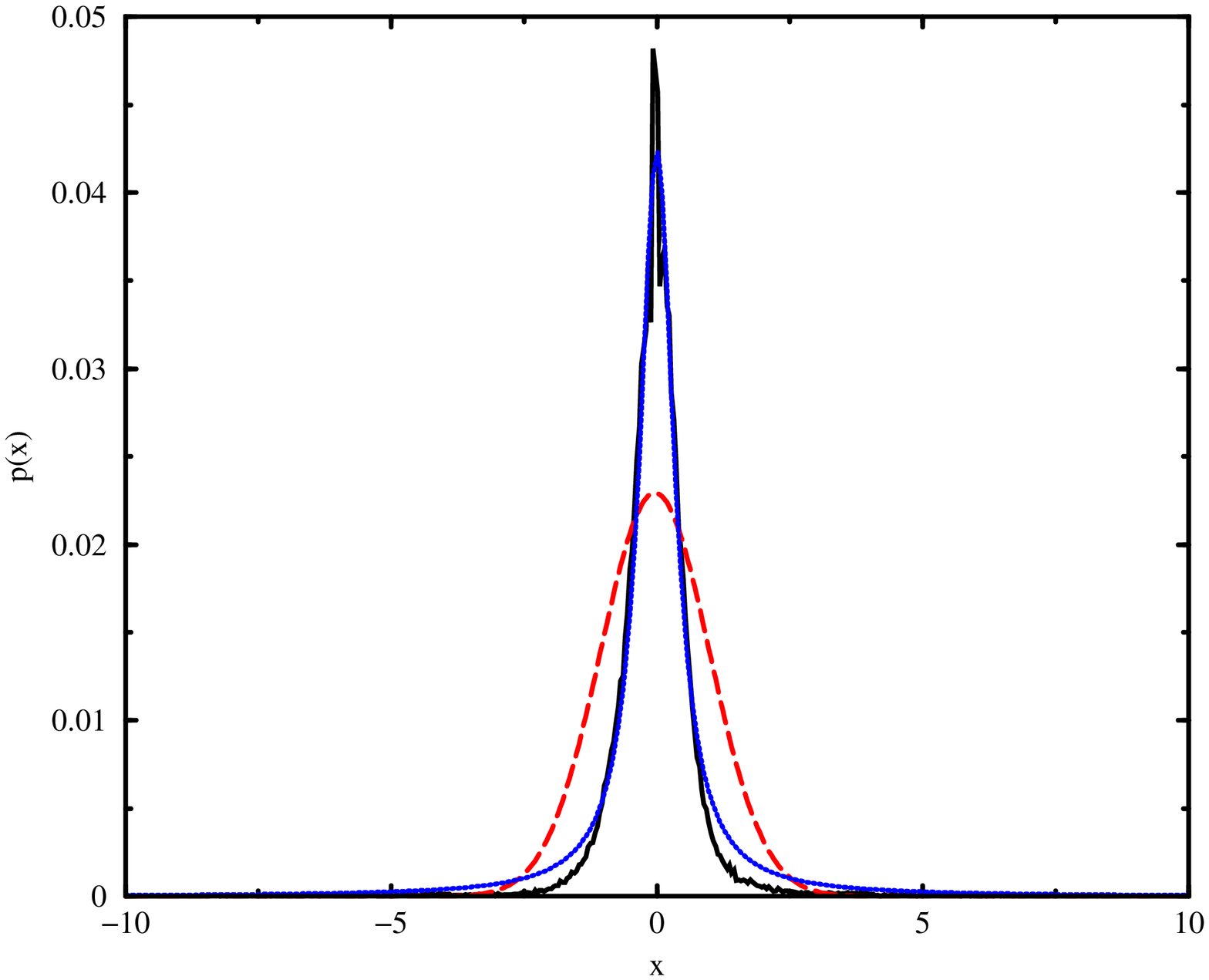}
\caption[0]{
Fitting of the probability distribution function
with the return time T=10min
by the Gaussian function (dashed line) and
by the Lorentzian (dotted line) function for
KOSDAQ.
}
\end{figure}

\begin{figure}
\includegraphics[width=7cm,angle=270]{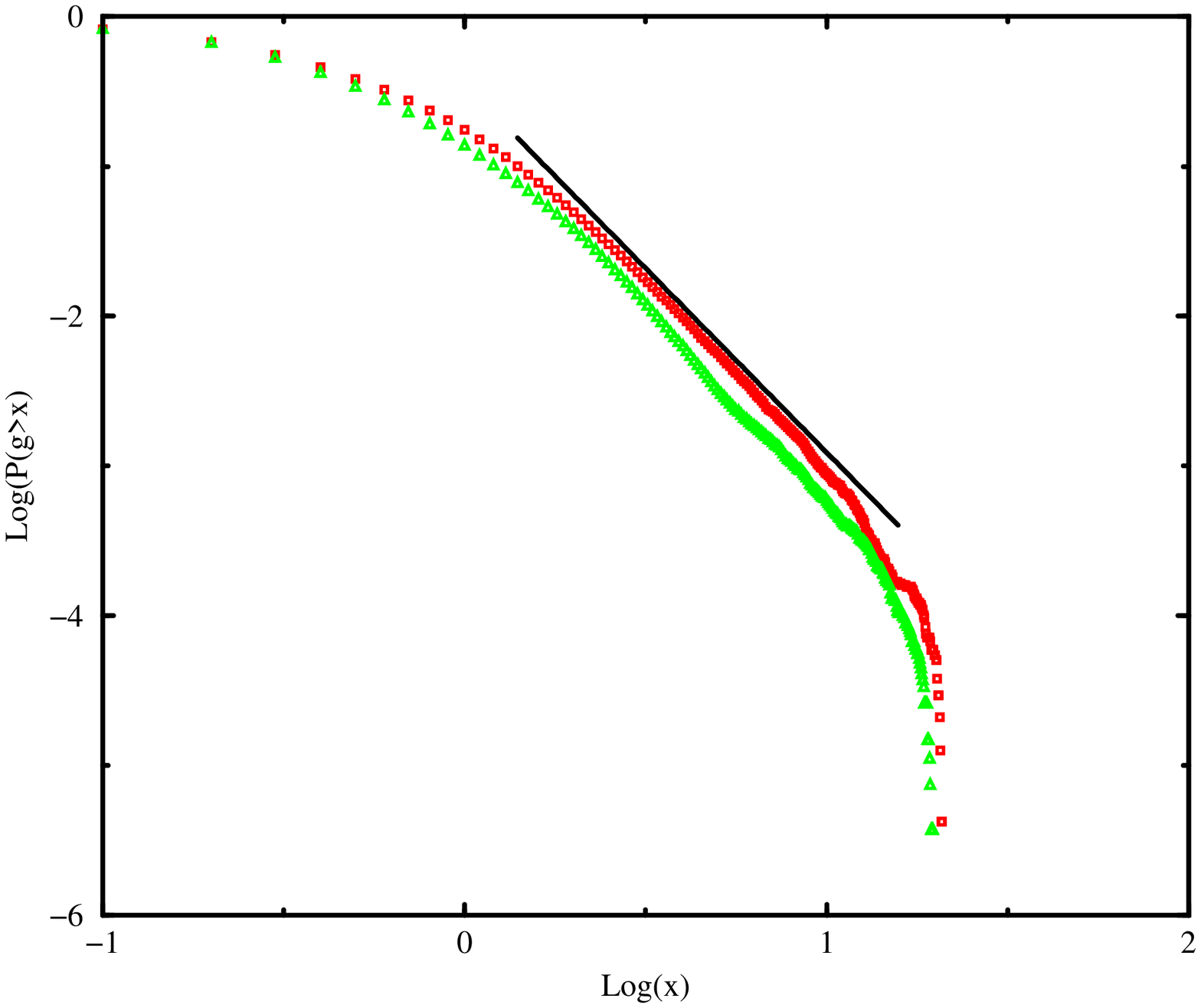}
\caption[0]{
The log-log plot of the accumulated probability distribution function
as a function of the return
with the return time T=10min for KOSPI.
}
\end{figure}
\begin{figure}
\includegraphics[width=7cm,angle=270]{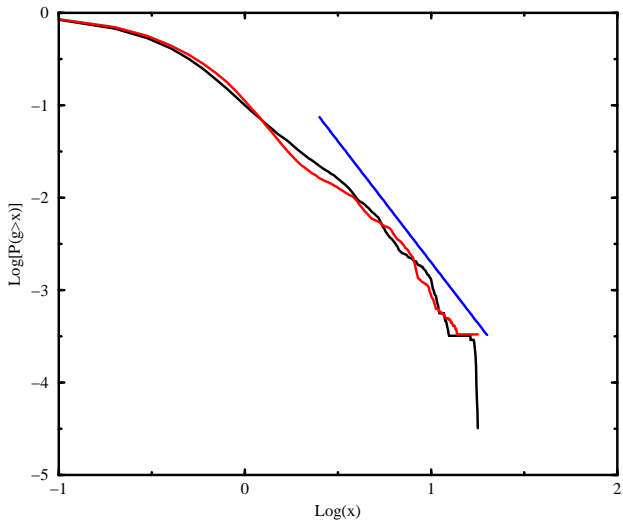}
\caption[0]{
The log-log plot of the accumulated probability distribution function
as a function of the return
with the return time T=10min for KOSDAQ.
}
\end{figure}

\section{RESULTS AND DISCUSSIONS}

The normalized return presented in Fig.1 for KOSPI and in Fig.2
for KOSDAQ. We observed large price changes around the period of
Asian economic crisis in September 1997. We consider the
logarithmic return with the return time T=1min, 10min, 30min,
60min, and 600min. The pdf for the return presented in Fig.3
(KOSPI) and Fig.4 (KOSDAQ).
 For the short return time T=1min, the pdf of the return has long tails with
very large fluctuations. The peak of the pdf of the return
decreases as the return time T increases. We fit the pdf with the
Gaussian distribution function and Lorentzian distribution
function for the return time T=10min in Fig.5(KOSPI) and
Fig.6(KOSDAQ). The central region of the pdf is fitted better by
Lorentzian than by Gaussian\cite{KIM}. However, the positive and
negative tail region of the pdf deviates from Gaussian and
Lorentzian.
 The tail of the pdf of returns decays according to a
power law as
\begin{equation}
p(x) \sim x^{-(1+\alpha)}
\end{equation}
with the exponent $\alpha > 2$. The accumulated pdf of returns
is defined as
\begin{equation}
P(g > x) = \int^{\infty}_{x} p(x) dx
\end{equation}
The accumulated pdf follows a power-law behavior as
\begin{equation}
P(g>x)  \sim \frac{1}{x^{\alpha}}
\end{equation}

In Fig.7(KOSPI)  and Fig.8(KOSDAQ), we presented the accumulated
probability distribution function for the return time T=10min. We
observed that the exponents $\alpha$ are greater than 2 which
means that the pdf of returns deviated from the stable L\'{e}vy
distribution with $0 < \alpha < 2$. We present the exponents
$\alpha$ for the different return time T in the table 1. The
exponent $\alpha$ was measured for different stock indices over
many countries. We summarized the measured exponent $\alpha$ in
the table 1 for many different stock indices. Exponents $\alpha$
of KOSPI and KOSDAQ increase when the return time increases. We
also observed that the range of the power-law diminish as the
return time increases. The pdf of the return deviated from the
stable L\'{e}vy distribution for all different stock indices. The
exponents $\alpha$ of the positive and negative tail are greater
than 2. The exponent $\alpha$ of the positive tail of KOSPI and
KOSDAQ are sightly less than one of the negative

%\begin{widetext}
%\begin{center}
\begin{table}
%\begin{table}
\caption[0]{Summary of exponents $\alpha$ for the pdf for the
positive tail}
\begin{ruledtabular}

\begin{tabular}{c|c|c|c} \hline
 & \multicolumn{2}{c}{positive tail} & \\ \hline
 & $\alpha$ & range & Ref\\ \hline
 & 2.16 (T=1min) & $1 < g < 12$ & [1]\\
 & 2.46 (T=10min) & $1.4 < g < 10$ & [1]\\
 KOSPI & 2.71 (T=30min) & $1.5 < g < 10$ & [1] \\
 & 2.87 (T=60min) & $2 < g < 7$ & [1] \\
 & 2.87 (T=600min) & $ 2  < g < 4$ & [1] \\ \hline
 & 2.06 (T=1min) & $10 < g < 30$ & [1] \\
 & 2.22 (T=10min) & $3 < g < 10$ & [1] \\
 KOSDAQ & 2.39 (T=30min) & $1.6 < g < 6.3$ & [1] \\
 & 2.70 (T=60min) & $1.6 < g < 4$ & [1] \\
 & 2.46 (T=600min) & $ 1  < g < 2.2$ & [1] \\ \hline
 & 2.4 (T=1min) & $2 <g < 20$ & [2] \\
 DAX & 2.9 (T=10min) &  & [2] \\
 & 3.5 (T=60min) &  & [2] \\
 & 3.5 (T=1day) &  & [2] \\
\hline
 & 2.32 (T=1min) & $ 3< g < 15 $ & [3] \\
Hang-Seng & 3.05 (T=1day) & $ 1 < g $  & [4] \\
 & 5.0 (T=1day) &  & [3] \\
\hline
 & 2.95 (T=1min) & $3 < g < 50 $ &  [4] \\
 & 3.45 (T=1min) &   & [4] \\
 & 2.69 (T=16min) &  & [4] \\
S \& P 500 & 2.53 (T=32min) &  & [4] \\
 & 2.83 (T=128min) &  & [4] \\
 & 3.39 (T=512min) &  & [4] \\
 & 3.34 (T=1day) &  & [4] \\
\hline
Nikkei & 3.05 (T=1day) & $ 1 < g $  & [4] \\
\hline
\end{tabular}

%\end{table}
\end{ruledtabular}
\end{table}

\begin{table}
%\begin{table}
\caption[0]{Summary of exponents $\alpha$ for the pdf for the
negative tail}
\begin{ruledtabular}

\begin{tabular}{c|c|c|c} \hline
 & \multicolumn{2}{c}{negative tail} & \\ \hline
 & $\alpha$ & range & Ref\\ \hline
 & 2.29 (T=1min) & $1 < g < 12 $ & [1]\\
 & 2.56 (T=10min) & $1.4 < g < 10 $ &  [1]\\
 KOSPI & 2.73 (T=30min) & $1.5 < g < 8 $ & [1] \\
 & 3.03 (T=60min) & $2 < g < 7 $ & [1] \\
 & 2.91 (T=600min) & $2 < g < 3.4$ & [1] \\ \hline
 & 2.41 (T=1min) & $10 < g < 30 $ & [1] \\
 & 2.62 (T=10min) & $3< g < 12 $ & [1] \\
 KOSDAQ & 1.89 (T=30min) & $1.6 < g < 6.3 $ & [1] \\
 & 2.09 (T=60min) & $1.6 < g < 6.3 $ & [1] \\
 & 1.88 (T=600min) & $1.2 < g < 2.5$ & [1] \\ \hline
 DAX & 2.6(T=1min) & $2<g<20$ & [2] \\
 \hline
 & 2.32 (T=1min) & $3< g < 15 $&  [3] \\
Hang-Seng
 & 4.0 (T=1day) &  & [3] \\
 \hline
 & 2.75 (T=1min) & $3 < g < 50$ & [4] \\
Hang-Seng
 & 3.29 (T=1min) &  Hill estimator & [4] \\
 \hline
\end{tabular}

%\end{table}
[1] Present work, [2] Ref.16, [3] Ref.17, [4] Ref.10
\end{ruledtabular}
\end{table}
%\end{center}
%\end{widetext}

From the analysis of the stock market indices, we observed that the pdf
of KOSPI index is well outside the stable L\'{e}vy distribution.
The exponent $\alpha$ depends on the return time $T$. The observed
exponent $\alpha$ increased when the return time increased for
both positive and negative tail. The pdf of KOSDAQ index also is outside
the stable L\'{e}vy distribution with $\alpha >2$ for the positive tail.

\section{CONCLUSIONS}

We consider the probability density function(pdf) of the Korean
stock market indices, KOSPI and KOSDAQ indices. We observe that
the pdf for both  KOSPI and KOSDAQ indices fit neither Gaussian
nor Lorentzian distribution function. Central parts of indices are
well fitted by the Lorentzian distribution function. However, the
tail parts of the pdf deviate strongly form Gaussian and
Lorentzian distribution function. The tail part of the pdf follows
a power-law asymptotic behavior. We observe that exponents of the
power-law of the accumulated probability distribution function are
well outside the stable L\'{e}vy distribution. Korean stock market
is not described by the random Gaussian stochastic processes.

\begin{acknowledgements}
This work was supported by KOSEF(R05-2003-000-10520-0(2003)).
\end{acknowledgements}

\newcommand{\jpa}{J. Phys. A}
\newcommand{\jkps}{J. Kor. Phys. Soc.}


\begin{thebibliography}{}

 \bibitem{MA99} R.N. Mantegna and H.E. Stanley, {\em An Introduction to
Econophysics: Correlations and Complexity in Finance}, Cambridge University Press,
Cambridge, 1999.
 \bibitem{MA97} B. Mandelbrot, {\em Fractals and Scaling in Finance}, Springer,
New York, 1997.
 \bibitem{MS95} R.N. Mantegna, H.E. Stanley, Nature, 376(1995) 46.
 \bibitem{BS94} J.-P. Bouchaud, D. Sornette, J. Phys. I France 4(1994) 863.
 \bibitem{BA1900} L. Bachelier, Ann. Sci. \'{E}cole Norm. Sup. 3(1900), 21.
 \bibitem{MA63} B. Mandelbrot, J. Business, 36(1963) 294.
 \bibitem{FA63} E.F. Farma, J. Business, 36(1963) 420.
 \bibitem{LE} P. L\'{e}vy, {\em Theorie de l'addition des variables al\'{e}atories},
Gauthier-Villars, Paris, 1934.
 \bibitem{PA} V. Pareto, {\em Cors d'Economie Politique}, Lausanne, Paris, 1897.
 \bibitem{GP99} P. Gopikrishnan, V. Plerou, L.A.N. Amaral, M. Meyer, H.E. Stanley, Phys.
Rev. E 60(5305) 1999.
 \bibitem{GM99} P. Gopikrishan, M. Meyer, L.A.N. Amaral, H.E. Stanley, Eur. Phys. J. B.
3(1999) 139.
 \bibitem{LK99} Y. Liu, P. Gopikrishnan, P. Cizeau, M. Meyer, C.-K. Peng, H.E. Stanley,
Phys. Rev. E 60(1999), 1390.
 \bibitem{GP00} P. Gopikrishnan, V. Plerou, Y. Liu, L.A.N. Amaral, X. Gabaix, H.E. Stanley,
Physica A 287(2000), 362.
 \bibitem{SA00} H.E. Stanley, L.A.N. Amaral, P. Gopikrishnan, V. Plerou, Physica A
283(2000), 31.
\bibitem{LH03} J. W. Lee and B. H. Hong, \jkps 43, 303(2003).
\bibitem{CL01} S. H. Cheon and J. W. Lee, \jkps 38, 782(2001).
\bibitem{LL02} K. E. Lee and J. W. Lee, \jkps 40, 385(2002).
\bibitem{KC01} Y. Kim, S. H. Choi, and S. Y. Yoon, \jkps 38, 500(2001).
\bibitem{KP03}  S. J. Koo, S. Y. Park, and J. W. Lee, \jkps 42, 331(2003).
\bibitem{KK02} H.-J. Kim, I.-M. Kim, and B. Khang, \jkps 40, 1105(2002).
\bibitem{NA02} Y. Nakajima, \jkps 40, 1096(2002).
 \bibitem{EV95} C.J.G. Evertsz, Proceedings of the First International Conference
on High Frequency Data in Finance, Z\"{u}lich, 1995.
\bibitem{GD02} A.Z. G\'{o}rski, S. Drodz, J. Septh, Physica A 316(2002), 496.
\bibitem{WH01} B.H. Wang, P.M. Hui, Eur. Phys. J. B. 20(2001) 573.
\bibitem{KIM} K. Kim, S.-M. Yoon, cond-mat/0305270(2003).
% Text of bibliographic item
\end{thebibliography}
\end{document}